\documentclass[aps,pra,reprint,superscriptaddress,nofootinbib]{revtex4-2}

\usepackage{amsmath,amssymb,amsthm}
\usepackage{booktabs}
\usepackage{graphicx}
\usepackage{xcolor}
\usepackage{hyperref}
\usepackage{microtype}

\usepackage{algorithm}
\usepackage{algpseudocode}

\hypersetup{
  pdftitle={A Path to Practical Fault-Tolerance Through Large-scale Hyperbolic Codes},
  pdfauthor={Ahmed Adel Mahmoud and Steven Rayan},
  colorlinks=true,
  linkcolor=blue!55!black,
  citecolor=blue!55!black,
  urlcolor=blue!55!black
}

\begin{document}
\raggedbottom

\title{Geometry-optimized hyperbolic codes for modular fault-tolerant quantum architectures}

\author{Ahmed Adel Mahmoud}
\email{ahmed.mahmoud@usask.ca}
\affiliation{Centre for Quantum Topology and Its Applications (quanTA), University of Saskatchewan, Saskatoon, Saskatchewan, Canada}
\affiliation{Department of Mathematics and Statistics, University of Saskatchewan, Saskatoon, Saskatchewan, Canada}

\author{Steven Rayan}
\email{rayan@math.usask.ca}
\affiliation{Centre for Quantum Topology and Its Applications (quanTA), University of Saskatchewan, Saskatoon, Saskatchewan, Canada}
\affiliation{Department of Mathematics and Statistics, University of Saskatchewan, Saskatoon, Saskatchewan, Canada}

\date{\today}

\begin{abstract}
Quantum fault tolerance is a prerequisite to harness the full potential of quantum computers. 
However, achieving fault tolerance remains a major challenge since it requires optimizing tradeoffs between several intertwined theoretical and experimental parameters, including the encoding rate, code distance, parity-check weights, code planarity, and routing.
Surface codes remain strong candidates for fault tolerance due to their simple geometric structure, established decoding methods, locality, and experimental attainability.
In this work, we present finite families of hyperbolic surface codes that leverage the optimal efficiency scaling $kd^2/n = C (\log{k})^2$ attainable by surface codes with bounded local geometry.
We show that optimizing the periodic identifications can double the code distance and thereby increase the code's efficiency by a factor of four, at fixed qubit count, encoding rate, and parity-check weight.
We also address the routing problem of hyperbolic surface codes by introducing a topology-aware algorithm that partitions a hyperbolic code into bounded planar modules while reducing long-range communications demand.
Circuit-level Monte Carlo simulations yield finite-size threshold estimates for both modular and monolithic layouts, showing that tripling the long-range gate error probability mildly lowers, but does not eliminate, the estimated error-correction thresholds.
Together, these results strengthen the case for hyperbolic surface codes as structured and highly efficient large-scale quantum error correction codes for future fault-tolerant architectures.

\end{abstract}

\maketitle

\section{Introduction}

Quantum error correction is essential for protecting quantum information against imperfect operations and environmental noise. A central challenge is to achieve this protection with resources that remain manageable for large-scale quantum computers \cite{gottesman1998theory,gottesman2010introduction}.
Quantum low-density parity-check (qLDPC) codes offer a promising route to fault tolerance as they ensure that parity-check weights and individual qubit degrees remain bounded independently of the overall code size \cite{BreuckmannEberhardt2021}. 
Many qLDPC codes have constant encoding rates and code distance growing with the system size \cite{BreuckmannEberhardt2021, Terhal2015}. 
Realizing the potential of qLDPC codes, however, requires balancing encoding rate, code distance, check weight, and the physical interactions required to perform syndrome measurements.

However, several theoretical bounds restrict qLDPC codes parameters and control the choices available to design new codes. For a surface code with parameters \([[n,k,d]]\), there is a tradeoff between the encoding rate $k/n$ and the square of the code distance $d^2$. For instance, for codes with bounded interaction range on a two-dimensional Euclidean lattice of fixed qubit density, the Bravyi--Poulin--Terhal (BPT) bound gives \(kd^2/n=O(1)\) \cite{BravyiPoulinTerhal2010}. Thus, a growing distance requires a vanishing encoding rate \(k/n\ \rightarrow 0\) under these assumptions. Conventional Euclidean surface codes obey this bound, but remain attractive blueprints for quantum fault tolerance due to their simple geometric structure \cite{Kitaev2003,Dennis2002,Terhal2015}, established decoding methods \cite{fowler2012towards, delfosse2021almost, bausch2024learning}, developed protocols for logical operations \cite{horsman2012surface, bravyi2005universal, fowler2012surface}, as well as their experimental accessibility \cite{google2023suppressing, krinner2022realizing, zhao2022realization, google2025quantum} (see also Ref.~\cite{Rayan2026} for a 
generalized field-theoretic perspective on topological surface codes). At the same time, their resource overhead motivates the search for more efficient alternatives, with similar geometric, decoding and experimental attainability advantages.

Relaxing geometric locality creates new possibilities, but sparse checks alone do not guarantee favorable code parameters. Baspin and Krishna relate the achievable distance and encoding rate to how readily the qubit connectivity graph can be separated into smaller pieces \cite{BaspinKrishna2022,BaspinEtAl2024}. More concretely, they prove that for any qLDPC code, the code distance is upper bounded by the separation profile of the code's connectivity graph. 

Check weight imposes further restrictions. Stabilizer codes with checks of weight at most three cannot have distance greater than two if they encode any logical information \cite{WangEtAl2026}. For CSS codes, whose checks separate into \(X\)-type and \(Z\)-type operators, a weight-four code in which every qubit belongs to exactly two checks of each type obeys \(kd^2/n=O(1)\), even without assuming Euclidean locality \cite{WangEtAl2026}. Within this structure, exceeding constant efficiency \(kd^2/n=O(1)\) therefore requires checks of weight at least five.
These bounds motivate constructions that retain modest check weights while maintaining rich global connectivity.

Tile codes relax parity-check locality to improve upon conventional surface-code patches while retaining a planar lattice construction~\cite{tile_codes_prl,BreuckmannEtAl2025TileLogical}. 
Concurrently, bivariate-bicycle (BB) and barbell codes leverage modular architectures to achieve highly competitive encoding rates and distances~\cite{BravyiEtAl2024,eberhardt2024logical,eberhardt2024pruning,choe2026barbell}. 
Although their favorable finite-size scaling is highly promising for near-term devices, it does not by itself guarantee advantageous asymptotic scaling. 
Specifically, bounded-range planar tile and barbell families remain fundamentally limited by the BPT bound, $kd^2/n = O(1)$. 
For BB codes, two-layer routing alone does not establish uniformly bounded geometric range~\cite{BravyiEtAl2024}. BPT applicability must therefore be assessed for the particular BB family; any family with bounded stabilizer range on a two-dimensional lattice, even with periodic boundaries, obeys the same bound.
A distinct scaling limitation arises in the Pinnacle architecture, which employs a specific sequence of generalized-bicycle codes~\cite{WebsterEtAl2026Pinnacle,WebsterEtAl2025Gadgets}. For these codes, the parameters $n=2(2^m-1)$ and $k=2m$, combined with the conjectured distance $d=m+(m-4)^2$, yield a vanishing asymptotic efficiency scaling of $kd^2/n = \Theta((\log n)^5/n) \to 0$. 
While replicating fixed-size code blocks successfully preserves the block-wise encoding rate \cite{BreuckmannEberhardt2021}, it inherently fails to scale the code distance. Consequently, these examples underscore a critical challenge in quantum error correction: characterizing how code efficiency $kd^2/n$ scales as the code size increases beyond small code instances, while accounting for the physical routing and operational overheads of large-scale fault-tolerant architectures.

Other qLDPC codes have been shown to achieve the stronger scaling, with both the number of logical qubits and the distance growing proportionally to the physical-qubit count \cite{panteleev2022asymptotically,LeverrierZemor2022}. Such families are called asymptotically good, but usually feature high-weight parity checks \cite{lin2022good}.
Weight-reduction methods can lower pariity-check weights and qubit degrees while preserving asymptotic goodness, at the cost of additional physical qubits~\cite{WangEtAl2026,HastingsWeightReduction}.
Nevertheless, their practical value depends on the resource overheads associated with syndrome extraction circuits, physical qubit routing, and the execution of fault-tolerant logical operations.

On the other hand, hyperbolic surface codes preserve the surface-based description of checks and logical operators, while offering a compelling combination of encoding efficiency, growing distance, and geometric structure \cite{freedman2002z2, albuquerque2009topological, BreuckmannTerhal2016}. They are defined on closed higher-genus Riemann surfaces whose number of handles grows with the code size, allowing $k$ to grow proportionally to \(n\). For surface codes with bounded face sizes and vertex degrees, Delfosse established the upper bound
\[
\frac{kd^2}{n}\leq C(\log k)^2,
\]
where \(C\) depends on the local geometry \cite{Delfosse2013}. This permits the code's efficiency $kd^2/n$ to grow with code size beyond the constant bound imposed by Euclidean locality. The bound alone does not guarantee that a particular construction realizes this growth. A central question, that we address in this work, is therefore how closely explicit finite surface codes can approach this scaling while retaining low-weight checks and manageable connectivity.

Translating this theoretical potential into a viable superconducting architecture requires bridging abstract hyperbolic geometry with explicit stabilizer codes and physical hardware constraints. In particular, realizing hyperbolic surface codes introduces four fundamental hurdles: (i) systematically constructing logical operators that satisfy the canonical Pauli commutation relations, (ii) embedding non-Euclidean negative curvature into a planar two-dimensional superconducting layout, (iii) optimizing periodic identifications to obtain code families that leverage the optimal efficiency scaling \cite{Delfosse2013}, and (iv) mitigating the extensive non-local routing required by the periodic lattice connectivity.
Our earlier work addressed the first challenge by establishing a systematic framework for identifying parity checks and logical operators for hyperbolic codes \cite{MahmoudAliRayan2026}, while our subsequent work addressed the second by engineering capacitive couplings between superconducting resonators to emulate the Poincaré metric \cite{XuEtAl2025}. While this demonstration provided a viable physical platform for open-boundary hyperbolic connectivity, realizing these codes as fault-tolerant devices requires scalable routing protocols to implement long-range connectivity. In this work, we address the third and fourth challenges by demonstrating that substantial efficiency gains emerge from optimizing the compactification of the hyperbolic surface, in addition to introducing a topology-aware compiler to modularize hyperbolic codes, while systematically assessing the routing overhead in both monolithic and modular architectures.

\begin{figure*}[t]
\centering
\includegraphics[width=\textwidth]{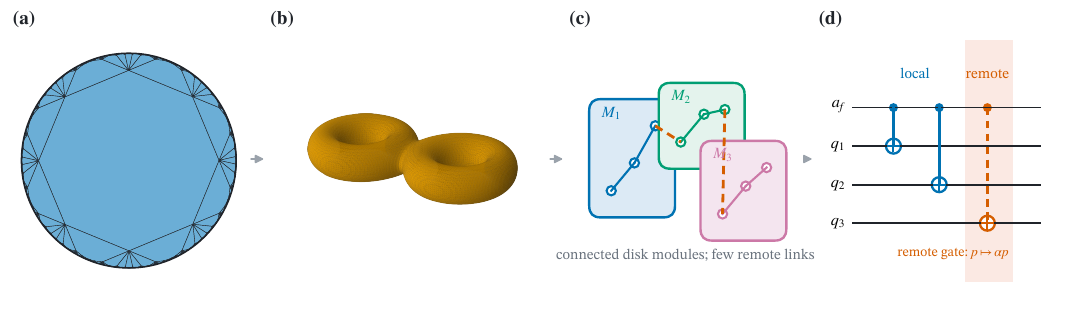}
\caption{From code geometry to hardware implementation. (a) A regular \(\{8,8\}\) tiling in the Poincar\'e disk model, illustrating the self-dual tilings studied here. (b) A schematic closed genus-two surface illustrating imposing periodic boundary conditions; the computed codes generally have much larger genus. (c) A modular realization with connected planar regions and inter-module links. (d) A measurement circuit distinguishing local and remote gates. Assigning qubits to modules changes the physical organization of interactions while preserving the hyperbolic code's parameters.}
\label{fig1:overview}
\label{fig:overview}
\end{figure*}

Specifically, we perform an exhaustive search over self-dual prime-field  $\{p,p\}$ tessellations for $p \in \{5,6,7,8\}$, uncovering code efficiency of $kd^2/n \simeq 33.34$ in a $[[51330, 17112, 10]]$ code that harbors over $17,000$ logical qubits. We demonstrate that tailoring the periodic boundary conditions can double the code distance at fixed qubit count, encoding rate, and check weight. Furthermore, selected $\{5,5\}$ and $\{7,7\}$ code families exhibit finite-size scaling consistent with the optimal code-efficiency scaling allowed by Delfosse's bound. 
To facilitate hardware implementation, we introduce a topology-aware modular partitioning algorithm that preserves the code structure and satisfies the prescribed capacity and geometric constraints.
Its reference layouts reduce all five communication metrics relative to the tested load-feasible spectral-partitioning assignments.
Finally, via circuit-level Monte Carlo simulations, we quantify threshold degradation under noisier long-range gates across both monolithic and modular architectures (summarized in Fig.~\ref{fig1:overview}). We envision that, alongside our preceding results, this work establishes a concrete architectural blueprint toward scalable, high-rate fault-tolerant quantum memories based on hyperbolic surface codes.

Section~\ref{sec:construction} presents the code construction, search, and distance calculations. Section~\ref{sec:family} examines finite-size efficiency scaling. Sections~\ref{sec:architectures} and~\ref{sec:compiler} address the physical-connectivity and introduces our modular partitioning algorithm, respectively. Section~\ref{sec:circuit} reports memory performance, while Sec.~\ref{sec:discussion} concludes with the implications and explores future directions and requirements for practical fault-tolerant architectures based on hyperbolic surface codes.

\section{Code construction, exhaustive search, and distance calculation}
\label{sec:construction}

A regular hyperbolic tiling specifies the local structure of the code, while boundary identifications compactify a finite region into a closed surface and determine its global connectivity. For a \(\{p,p\}\) tiling, we utilize the orientation-preserving triangle group
\begin{equation}
\Delta_p=\langle x,y\mid x^p=y^p=(xy)^2=1\rangle,
\label{eq:triangle}
\end{equation}
to define the underlying hyperbolic tiling. A finite quotient
\begin{equation}
Q=\Delta_p/\Gamma_{\mathrm{PBC}}
\end{equation}
then specifies the periodic identifications. Faces, vertices, and edges are represented by cosets of the subgroups \(\langle x\rangle\), \(\langle y\rangle\), and \(\langle xy\rangle\), respectively \cite{Siran2001,MahmoudAliRayan2026}.

Data qubits occupy the edges, while each face defines a \(Z\) check on its boundary edges, and each vertex defines an \(X\) check on its incident edges. These checks commute because a face boundary meets each vertex an even number of times. For a closed, connected, orientable surface of genus \(g\), the code encodes \(k=2g\) logical qubits.

The local tiling structure fixes several code parameters before the periodic identifications are chosen. Writing \(F\), \(V\), and \(E\) for the numbers of faces, vertices, and edges, respectively, gives
\begin{equation}
pF=pV=2E,\qquad n=E.
\end{equation}
Euler's relation then yields
\begin{equation}
\frac{k}{n}=\frac{p-4}{p}+\frac{2}{n}.
\label{eq:rate}
\end{equation}
Both check types have weight \(p\), and each data qubit participates in two checks of each type. Assigning one ancilla qubit to every check requires \(V+F=4n/p\) ancilla qubits and \(4n\) data--ancilla interactions per syndrome round.

At fixed tiling and quotient-group order, altering the periodic identifications leaves these parameters unchanged but can alter the shortest logical operators. Therefore, periodic identifications provide a means of improving the code distance at fixed physical-qubit count, encoding rate, and check weight.

We restrict the quotient group to $Q \cong \operatorname{PSL}(2,q)$, where $q$ is prime. This group comprises $2 \times 2$ determinant-one matrices over the finite field $\mathbb{F}_q$, taken modulo its center. For odd $q$, the group order is $|Q| = q(q^2-1)/2$. Because the data qubits of the code reside on the edges of the tessellation—which correspond algebraically to the cosets of the order-two subgroup $\langle xy \rangle$—the total number of data qubits is exactly $n = |Q|/2$. 
Here, $q$ determines the quotient-group order and hence the code size, while $p$ determines the local tiling; admissible pairs must satisfy the generator conditions below.

To enforce the lattice's self-duality, we introduce a projective involution $J \in \operatorname{PGL}(2,q)$ and constrain the generators such that $y = JxJ^{-1}$. We consider both split and non-split involutions, characterized respectively by having two or zero fixed points on the projective line over $\mathbb{F}_q$. Valid generating pairs $(x,y)$ must satisfy the triangle-group relations $\operatorname{ord}(x) = \operatorname{ord}(y) = p$ and $\operatorname{ord}(xy) = 2$, while collectively generating the full quotient group $Q$. By construction, conjugation by $J$ swaps $x$ and $y$, structurally embedding a strict face--vertex symmetry into the resulting code.

To remove redundancies under the chosen algebraic equivalence, we group generator pairs into orbits under conjugation by the centralizer of $J$ in $\operatorname{PGL}(2,q)$ (the subgroup of elements that commute with $J$).
We retain only a single representative from each orbit, and the class counts reported below refer to this specific equivalence relation.
We perform an exhaustive enumeration of this algebraic construction
for tiling parameters $p\in\{5,6,7,8\}$ across all 17 primes $q<60$, giving 68 tiling--field pairs. Recording both swap labels for each
pair gives 136 cases, of which 115 contain no candidate generators.
The remaining 21 cases produce 2,304 candidate generator pairs,
which form 38 centralizer classes.
Each retained class undergoes verification to ensure it satisfies the triangle-group relations, generates the correct group order, and forms a valid, closed, orientable, and fully connected cellulation. 
The exhaustive claim applies specifically to regular, closed $\{p,p\}$ hyperbolic codes that admit the specified projective face--vertex swap and possess prime-field $\operatorname{PSL}(2,q)$ quotient groups.

For these hyperbolic surface codes, logical operators correspond to non-trivial homological cycles on the code manifold: $Z$-type logical operators map to cycles in the primal graph, while $X$-type logical operators map to cycles in the dual graph. A cycle is considered logically trivial (a stabilizer) if its edge vector can be expressed as a linear combination of face boundaries over $\mathbb{F}_2$. The homological code distance is dictated by the shortest non-trivial cycle, commonly referred to as the systole of the surface, such that $d = \min(d_X, d_Z)$. Because the primal and dual graphs can theoretically exhibit different systole lengths, the code distance is computed independently to determine both $d_X$ and $d_Z$.

To systematically evaluate and compare these codes, we adopt the efficiency metric 
\begin{equation}
\eta=\frac{kd^2}{n}.\label{eq:efficiency}
\end{equation}
Table~\ref{tab:optimal-codes} presents the maximum-efficiency representative for each tiling parameter derived from the exhaustive search. The global optimum within this parameter space is achieved by the \(\{6,6\}\) code \([[51330,17112,10]]\), which exhibits a remarkable encoding efficiency of \(\eta\simeq33.34\).

\begin{table}[h!]
\caption{Maximum-efficiency representatives for each tiling parameter obtained from the exhaustive prime-field search up to \(q < 60\). The asymptotic encoding rate is denoted by \(r_\infty=(p-4)/p\), with both \(X\)- and \(Z\)-type checks possessing uniform weight \(w=p\).
Note that optimality is restricted to the algebraic construction outlined in this section.}
\label{tab:optimal-codes}
\centering
\small
\setlength{\tabcolsep}{8pt}
\begin{tabular}{@{}ccccc@{}}
\toprule
\(p=w\) & \(r_\infty\)  & \(q\) & \([[n,k,d]]\) & \(\eta\) \\
\midrule
5 & \(1/5\) &  59 & \([[51330,10268,12]]\) & 28.81 \\
6 & \(1/3\) &  59 & \([[51330,17112,10]]\) & 33.34 \\
7&\(3/7\) &41& \([[17220,7382,8]]\)&27.44\\
8 & \(1/2\) &  47 & \([[25944,12974,8]]\) & 32.00 \\
\bottomrule
\end{tabular}
\end{table}

This comparison underscores why the encoding rate alone is an insufficient metric for overall code performance. 
Although the \(\{8,8\}\) construction boasts the highest asymptotic rate, the superior distance of the optimal \(\{6,6\}\) code ultimately yields a higher overall efficiency. 
Conversely, the \(\{5,5\}\) construction achieves a maximum distance of 12 using weight-five checks, albeit bounded by a lower asymptotic rate of one fifth.

\begin{figure*}[t]
\centering
\includegraphics[width=\textwidth]{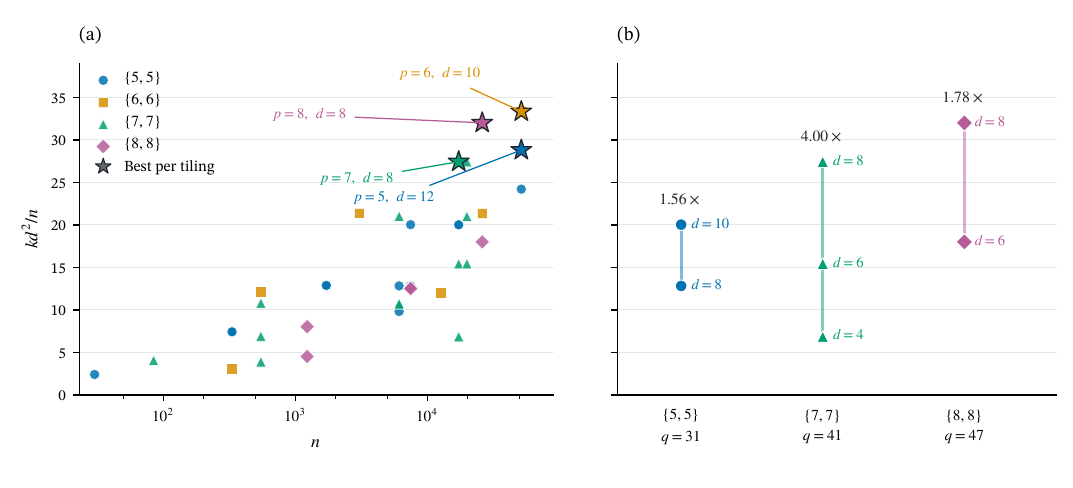}
\caption{Code efficiency and the role of periodic identifications. (a) Code efficiency \(kd^2/n\) versus physical-qubit count \(n\) for all 38 classes in the exhaustive search over \(p=5,6,7,8\) and prime \(q<60\), within the specified regular generator-swap-self-dual \(\operatorname{PSL}(2,q)\) construction. Colors and marker shapes distinguish the tilings; stars identify the maximum-efficiency code for each tiling. Classes with identical plotted parameters overlap. (b) Largest best-to-worst efficiency ratios at fixed tiling and quotient size within each family exhibiting variation in the searched class. Within each group, \(n\), \(k\), and check weight remain fixed. Point labels give distances; factors give \(\eta_{\max}/\eta_{\min}=(d_{\max}/d_{\min})^2\). 
For \(\{8,8\}\), the advantage of \(q=17\) ties \(q=47\); the latter contains the maximum-efficiency representative. 
The \(\{6,6\}\) family has one class per admissible \(q\) and no observed within-group variation.}
\label{fig:systematic-search}
\end{figure*}

Figure~\ref{fig:systematic-search} isolates the impact of periodic identifications by comparing distinct generating maps that share identical tiling parameters, quotient groups, physical-qubit counts, check-weights and encoding rates. For instance, within the \(\{7,7\}\) construction at \(q=41\), different identifications yield exact distances of 4, 6, and 8. Selecting the distance-eight map over its distance-four counterpart doubles the code distance, thereby quadrupling the efficiency \(\eta\). Crucially, this substantial enhancement is achieved entirely through the optimization of periodic boundary conditions; it incurs no additional overhead in data-qubit count, hardware check weight, or the number of two-qubit interactions required per syndrome extraction round.

To place these results in context, we compare our static constructions against the recently developed hyperbolic and semi-hyperbolic Floquet codes~\cite{HiggottBreuckmann2024}. While the static surface codes studied here require the direct measurement of fixed multi-qubit stabilizer generators, Floquet codes dynamically maintain the code space through a periodic schedule of two-qubit measurements. Nevertheless, the effective logical parameters of both paradigms are fundamentally governed by the topology of the underlying hyperbolic manifold, allowing for a direct comparison of their code-parameter efficiencies. Here, $d$ represents the homological distance, while for the Floquet constructions, $d$ denotes the embedded homological distance defined in Ref.~\cite{HiggottBreuckmann2024}.

Among the hyperbolic Floquet instances reported in that work, the maximum code efficiency is $\eta \simeq 5.01$, realized by a $[[19200,962,10]]$ code~\cite{HiggottBreuckmann2024}. In contrast, within our enumerated regular self-dual $\operatorname{PSL}(2,q)$ constructions for prime $q < 60$, the maximum efficiency is attained by the $\{6,6\}$ code $[[51330,17112,10]]$. This construction achieves an efficiency of $\eta \simeq 33.34$, representing a roughly $\times 6.65$ improvement over the Floquet benchmark. 
Our optimal $\{7,7\}$ code yields an efficiency of $\eta \simeq 27.44$. This exceeds the baseline Floquet efficiency by a factor of approximately $5.48$ while utilizing roughly $10.3\%$ fewer data qubits than the $[[19200,962,10]]$ benchmark. The conventional Euclidean toric code leverages $\eta=1$; therefore, our best code instance achieves $\times33.34$ improvement over the toric code while retaining the surface-based geometric description.

\section{Finite-size efficiency scaling}
\label{sec:family}

The exhaustive search allows us to examine how code efficiency grows with the number of logical qubits. Delfosse's bound limits \(\eta=kd^2/n=\mathcal{O}((\log k)^2)\) for surface codes with bounded local geometry \cite{Delfosse2013}. We therefore compare the computed efficiencies with \((\log k)^2\) over the available finite sequences.

For each fixed tiling, we select the smallest physical-qubit count at which each new maximum distance is attained within the exhaustive search. This selection favors increasing protection without retaining larger codes that offer no distance gain.
The resulting \(\{5,5\}\) and \(\{7,7\}\) sequences are listed in Table~\ref{tab:scaling-sequences}. The \(\{5,5\}\) sequence uses weight-five checks and has rate
\begin{equation}
\frac{k}{n}=\frac{1}{5}+\frac{2}{n},
\end{equation}
while the \(\{7,7\}\) sequence uses weight-seven checks and has rate
\begin{equation}
\frac{k}{n}=\frac{3}{7}+\frac{2}{n}.
\end{equation}
The weight-five and weight-seven sequences span more than three and two orders of magnitude in logical-qubit count, respectively.

\begin{table}[t]
\caption{Parameters of the code sequences in Fig.~\ref{fig:family-scaling}. Each sequence retains the smallest code attaining each new maximum distance within its tiling in the prime-field search through \(q < 60\).}
\label{tab:scaling-sequences}
\centering
\small
\setlength{\tabcolsep}{7pt}
\begin{tabular}{@{}cccc@{}}
\toprule
Tiling & \(q\) & \([[n,k,d]]\) & \(\eta\) \\
\midrule
\(\{5,5\}\) & 5  & \([[30,8,3]]\)        & 2.40  \\
            & 11 & \([[330,68,6]]\)      & 7.42   \\
            & 19 & \([[1710,344,8]]\)    & 12.87 \\
            & 31 & \([[7440,1490,10]]\)  & 20.03  \\
            & 59 & \([[51330,10268,12]]\)& 28.81 \\
\midrule
\(\{7,7\}\) & 7  & \([[84,38,3]]\)       & 4.07   \\
            & 13 & \([[546,236,5]]\)     & 10.81 \\
            & 29 & \([[6090,2612,7]]\)   & 21.02 \\
            & 41 & \([[17220,7382,8]]\)  & 27.44\\
\bottomrule
\end{tabular}
\end{table}

\begin{figure*}[t]
\centering
\includegraphics[width=\textwidth]{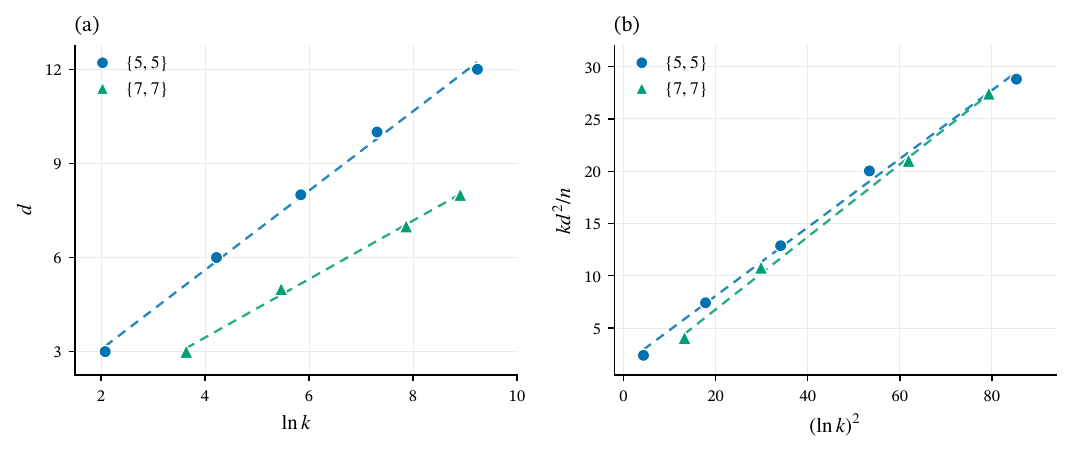}
\caption{Finite-size growth of the selected \(\{5,5\}\) and \(\{7,7\}\) code sequences from the prime-field search with \(q<60\). (a) Code distance \(d\) versus \(\ln k\). (b) Code efficiency \(kd^2/n\) versus \((\ln k)^2\). Each sequence retains the smallest code attaining a new maximum distance within its tiling and the searched class. Dashed lines show ordinary least-squares fits with intercepts, using the five \(\{5,5\}\) code instances and four \(\{7,7\}\) code instances presented in Table~\ref{tab:scaling-sequences}.}
\label{fig:family-scaling}
\end{figure*}

Figure~\ref{fig:family-scaling} compares the distance and code efficiency of the two selected sequences. The weight-five sequence increases the distance from 3 to 12 and the efficiency from 2.40 to 28.81. Its largest selected code encodes 10,268 logical qubits into 51,330 physical qubits. The weight-seven sequence increases the distance from 3 to 8 and the efficiency from 4.07 to 27.44, reaching 7,382 logical qubits. The largest selected \(\{5,5\}\) code offers smaller checks and greater distance, while the largest selected \(\{7,7\}\) code has a higher encoding rate.

To quantify the observed growth, we fit the distance as a linear function of \(\ln k\) and the efficiency as a linear function of \((\ln k)^2\). Ordinary least-squares fits to all points in each sequence give, for \(\{5,5\}\),
\begin{align}
d &\simeq 1.265\ln k+0.543,
& R_{\mathrm{fit}}^2&=0.997, \nonumber\\
\eta &\simeq 0.328(\ln k)^2+1.530,
& R_{\mathrm{fit}}^2&=0.996,
\label{eq:fit-55}
\end{align}
and, for \(\{7,7\}\),
\begin{align}
d &\simeq 0.932\ln k-0.278,
& R_{\mathrm{fit}}^2&=0.997, \nonumber\\
\eta &\simeq 0.347(\ln k)^2-0.148,
& R_{\mathrm{fit}}^2&=0.998.
\label{eq:fit-77}
\end{align}
Here, \(R_{\mathrm{fit}}^2\) is the coefficient of determination describing the agreement between each fitted curve and the data points.

The normalized code efficiency $\eta/(\ln k)^2$ ranges from approximately 0.338 to 0.555 for the $\{5,5\}$ sequence and from 0.308 to 0.362 for the $\{7,7\}$ sequence. 
The weight-five ratios decrease over the sampled sizes, whereas the weight-seven ratios vary
within a narrower interval. 
The observed growth is consistent with logarithmic-squared efficiency scaling over these finite
sequences.

Optimizing the periodic identifications is crucial in constructing these families.
Increasing $q$, and therefore the code size, does not always improve distance. 
The reported growth therefore describes the selected optimized sequences, rather than all codes obtained by increasing the code size.

These results establish explicit finite constructions with favorable efficiency growth at fixed check weight. Their immediate significance is that the observed advantage persists across large code instances required for large-scale fault-tolerant quantum computing. 
The following sections examine interaction layouts and circuit-level memory performance for these codes.

\section{From closed topology to physical connectivity}
\label{sec:architectures}
The same periodic identifications that create a high-rate code also determine long-range interactions that must be supplied by the hardware. Engineered couplings can mimic local connectivity of the hyperbolic metric, as demonstrated in the superconducting-resonator framework of Ref.~\cite{XuEtAl2025}. However, the long-range interactions augment chip design efforts with additional engineering challenges, such as devising routing protocols that minimize crosstalk between qubits.

Here we study two interaction layouts motivated by superconducting hardware: a modular processor that distributes connected planar regions across separate chips and uses inter-module gates to complete the checks, and a monolithic processor that places the code on one planar chip and supplies selected non-local gates through long-range routing channels. 
Existing work on networked error correction, modular qLDPC codes, and distributed hyperbolic Floquet memories motivates the modular layout~\cite{NickersonEtAl2014,StrikisBerent2023,SutcliffeEtAl2025,RametteEtAl2024}, alongside demonstrations of reconfigurable routers and gates between separate superconducting modules~\cite{WuEtAl2024,SongEtAl2025}. Cavity buses and long-range superconducting coupling schemes provide relevant building blocks for the monolithic layout~\cite{majer2007coupling,marxer2023long,xiong2025scalable,deng2025long}.

For both architectures, the communication overhead is determined by the two-qubit entangling gates executed between data and ancilla qubits. Because an edge connecting distant boundary vertices in the primal lattice does not generally correspond to a single non-local operation, we explicitly map both qubit species and enumerate the resulting long-range data--ancilla interactions. In the modular architecture, an interaction is considered remote if the participating qubits reside in distinct modules. In the monolithic architecture, an interaction is classified as long-range if realizing qubit adjacency requires routing across periodic boundary conditions.

In our circuit simulations, each required remote interaction is modeled as a two-qubit $\mathrm{CNOT}$ gate subject to an elevated error probability. 
In particular, a local $\mathrm{CNOT}$ gate operates with a baseline error probability $p$, whereas a remote gate incurs a proportional penalty, operating with an elevated error probability $\alpha p$ for some multiplier $\alpha \geq 1$.
This effective error model isolates the operational impact of remote gate infidelity under a fixed schedule, abstracting away physical link lengths and the lower-level overheads of routing buses, communication ports, or entanglement-generation protocols.

\section{Topology-aware modular compilation}
\label{sec:compiler}

In this section, we introduce a topology-aware compilation algorithm designed to partition closed hyperbolic surface codes into planar modules of bounded size. The primary objective is to minimize inter-module communication overhead while satisfying explicit geometric and hardware-capacity constraints.

The syndrome extraction circuit is modeled as a bipartite interaction graph, commonly referred to as a Tanner graph~\cite{tanner1981recursive}. The vertex set comprises the data qubits alongside one ancilla qubit for every face and vertex check; the edges represent the data--ancilla interactions. Consequently, for a regular \(\{p,p\}\) code, the total number of physical qubits (data plus ancillas) is \(n_{\mathrm{total}}=n(1+4/p)\). 

Given a module assignment mapping \(m(\cdot)\), we quantify the communication overhead using two primary metrics:
\begin{align}
R&=\bigl|\{(c,u):u\in S_c,\ m(c)\ne m(u)\}\bigr|,\label{eq:routing-R}\\
K&=\sum_c\left(\left|\{m(c)\}\cup\{m(u):u\in S_c\}\right|-1\right),\label{eq:routing-K}
\end{align}
where \(S_c\) denotes the data-qubit support of check \(c\). Here, \(R\) measures the total number of remote (inter-module) interactions per complete syndrome extraction round, while \(K\) quantifies the aggregate distribution of checks across multiple modules. Specifically, a check confined to a single module contributes zero to \(K\), whereas a check spanning three modules contributes two. The latter serves as an agnostic measure of check fragmentation, independent of the underlying physical routing protocol.

We also report three other communication metrics. The \textit{number of communicating module pairs} counts distinct pairs of modules connected by at least one data--ancilla interaction.
The \textit{maximum number of communication partners} is the largest number of distinct modules with which any one module interacts.
The \textit{maximum remote-interaction count} is the largest number of inter-module CNOTs involving any module during one syndrome-extraction round.
Each inter-module CNOT contributes one to the load of each participating module.

To model realistic hardware constraints, we impose a strict upper bound of \(B_{\max}=80\) qubits per module. The total number of modules $M$ is given by:
\begin{equation}
M=\left\lceil\frac{n_{\mathrm{total}}}{80}\right\rceil,\quad
\bar B=\frac{n_{\mathrm{total}}}{M},\quad
B_{\min}=\lfloor0.8\bar B\rfloor.
\label{eq:routing-load}
\end{equation}
Every module must encapsulate between \(B_{\min}\) and 80 qubits (including both data and ancillas). 
These capacity bounds are enforced during the qubit-placement phase and verified upon completion.
Furthermore, the local interaction graph of each module must remain both connected and planar. Topologically, the union of faces constituting a module must form a strict topological disk---a connected region featuring a single boundary loop and devoid of any handles.

Algorithm~\ref{alg:topology-compiler} details the topology-aware protocol, executed in three sequential phases: face partitioning, physical qubit assignment, and validation. We emphasize that this compiler acts solely as a physical routing map; it preserves the code structure as well as every data qubit and parity check, leaving the logical code parameters \([[n,k,d]]\) unaltered.

\begin{algorithm}[H]
\caption{Topology-aware modular compilation}
\label{alg:topology-compiler}
\small
\algrenewcommand\algorithmicrequire{\textbf{Input:}}
\algrenewcommand\algorithmicensure{\textbf{Output:}}
\begin{algorithmic}[1]
\Require $\{p,p\}$ hyperbolic surface code; bounds \(M, B_{\min}, B_{\max}\) from Eq.~\eqref{eq:routing-load}; maximum number of attempts \(T\).
\Ensure A partition into \(M\) face regions, a module assignment for every data and ancilla qubit, and the five communication metrics defined earlier. 
The assignment is accepted only if it satisfies all capacity and geometric constraints; otherwise, the run terminates as a failure.
\Statex \textbf{Step 1: Partition the faces}
\For{up to $T$ attempts}
    \State Initialize regions by selecting a random starting face, iteratively adding $M-1$ subsequent faces that maximize the dual-graph distance from the current set.
    \State Grow regions by adding at most one adjacent face per region, prioritizing smaller regions and candidates sharing more edges with the region; stop when complete or stalled.
    \State Reject any face addition that violates the topological disk condition or exceeds the maximum face limit \(f_{\max}\).
    \If{all faces are successfully assigned and every region contains at least \(f_{\min}\) faces}
        \State Retain the current face partition and \textbf{break}.
    \EndIf
\EndFor
\State \textbf{if} no valid face partition is found, \textbf{return} failure.
\Statex \textbf{Step 2: Assign data and ancilla qubits}
\State Data qubits whose two incident faces lie on one region are assigned to that region, while data qubits whose incident faces lie on two regions are assigned heuristically to either incident region.
\State Iteratively optimize data-qubit assignments by minimizing the cost function $\widetilde{C}$ in Eq.~\eqref{eq:routing-surrogate}. Terminate after 20 passes or when no accepted moves remain.
\State Assign each face-check ancilla to its corresponding face region.
\State Assign each vertex-check ancilla to a module containing at least one of its support qubits, minimizing remote interactions subject to the module capacity bounds \(B_{\min}\leq B_j\leq B_{\max}\).
\Statex \textbf{Step 3: Validate and measure}
\State \textbf{if} assignment fails to satisfy any capacity or geometric constraints \textbf{return} failure.
\State Compute the five communication metrics: \(R\), \(K\), the number of communicating module pairs, the maximum number of communication partners, and the maximum remote-interaction count.
\State \Return the face regions, qubit assignments, and communication metrics.
\end{algorithmic}
\end{algorithm}

The partitioning phase operates on the dual graph of the lattice, where vertices represent the faces of the primal lattice and edges connect adjacent faces. For a lattice with \(F\) total faces, the permissible number of faces per module is bounded by:
\begin{equation}
f_{\min}=\max\left\{2,\lfloor0.8F/M\rfloor\right\},\quad
f_{\max}=\lceil1.2F/M\rceil.
\end{equation}

With the face regions fixed, we optimize the permitted
data-qubit assignments using the Tanner graph.
Data-qubit optimization utilizes a cost function, \(\widetilde C\), formulated as:
\begin{equation}
\begin{aligned}
\widetilde C={}&\sum_c\bigl[10\bigl(|S_c|-\max_b t_{cb}\bigr)
+|\{b:t_{cb}>0\}|-1\bigr]\\
&+\sum_b[\max(0,\ell_b-L)]^2,
\end{aligned}
\label{eq:routing-surrogate}
\end{equation}
where \(t_{cb}\) denotes the number of data qubits belonging to check \(c\) that currently reside in module \(b\), \(\ell_b\) is the data-qubit load of module \(b\), and \(L=\min\{80,\lceil n/M\rceil+2\}\) is the target data-qubit capacity. 
A proposed data-qubit reassignment is accepted if it decreases
\(\widetilde C\), or leaves \(\widetilde C\) unchanged while decreasing
the load-penalty term.

Face-check ancillas are assigned to their corresponding face regions. The remaining vertex-check ancillas are assigned jointly to minimize remote interactions subject to module-capacity bounds, with each ancilla assigned to a module containing at least one of its support qubits. If module $j$ contains $b_j$ data qubits and face-check ancillas, it can receive between $\max(0,B_{\min}-b_j)$ and $B_{\max}-b_j$ vertex-check ancillas. The five communication metrics, described earlier, are evaluated only after all qubits have been placed.

To evaluate the compilation protocol, we benchmark one code for each check weight \(p \in \{5,6,7,8\}\). These instances are precisely the maximum-efficiency codes (\(kd^2/n\)) identified in the exhaustive search and presented in Table~\ref{tab:optimal-codes}. Consequently, all four instances are selected systematically by efficiency, independent of their innate routing performance. Notably, the \(\{5,5\}\) benchmark is the \([[51330,10268,12]]\) code and its reference module assignment is precisely the one utilized in the subsequent modular circuit simulations presented in Sec.~\ref{sec:circuit}. The reported reference assignments represent the initial recorded run for each code; we do not post-select the best communication cost across multiple trials.

We benchmark our topology-aware protocol against two standard graph-partitioning compilers. The first is recursive spectral bisection (RSB) applied to the dual graph, followed by the identical face-to-qubit assignment conversion used in our protocol. The second is Mt-KaHyPar~\cite{GottesbuerenEtAl2023}, applied directly to the Tanner graph to explicitly minimize the remote interaction count \(R\), utilizing its highest-quality single-thread configuration. Spectral partitioning is a well-established baseline, having been employed in previous distributed hyperbolic-code studies~\cite{SutcliffeEtAl2025,Galimova2026}. In our benchmarking, all methods enforce the same module count \(M\) and upper capacity of 80 qubits. During ancilla placement, RSB enforces the same lower capacity bound, whereas Mt-KaHyPar enforces only the upper bound natively, with lower bounds verified \textit{post hoc}. Crucially, neither baseline enforces the geometric constraints required for modular planarity; these topological properties are evaluated independently of the communication cost.

\begin{table*}[t]
\caption{Four benchmarks of the topology-aware compiler using the highest-efficiency codes presented in Table~\ref{tab:optimal-codes}. Here, \(n_{\mathrm{total}}\) counts data and ancilla qubits, and \(M\) is the module count.
The module-load interval and the values \(R\) and \(K\) describe the reference topology-aware layout for each code.
The medians \(R_{\mathrm{median}}\) and \(K_{\mathrm{median}}\) and relative ranges \(\delta_R\) and \(\delta_K\), defined in
Eq.~\eqref{eq:routing-variation}, summarize five independent initializations per code, including the initial reference layout.
All 20 layouts satisfy the load and geometric constraints.}
\label{tab:routing-instances}
\centering\small
\begin{tabular}{ccccccccccc}
\toprule
Tiling & \([[n,k,d]]\) & \(n_{\mathrm{total}}\) & \(M\) & Module load & \(R\) & \(K\)   & \(R_{\textrm{median}}\) & \(K_{\textrm{median}}\) & \(\delta_R\) (\%) & \(\delta_K\) (\%)    \\
\midrule
\(\{5,5\}\)&\([[51330,10268,12]]\)&\(92,394\)&\(1,155\)&77--80&\(69,021\)&\(57,382\)  &\(69,012\) &\(57,382\) & 0.60 & 0.48 \\
\(\{6,6\}\)&\([[51330,17112,10]]\)&\(85,550\)&\(1,070\)&68--80&\(83,962\)&\(69,544\)  &\(84,068\) & \(69,630\) & 0.28 & 0.21 \\
\(\{7,7\}\)&\([[17220,7382,8]]\)&\(27,060\)&\(339\)&67--80&\(31,967\)&\(25,723\)  &\(31,939\) & \(25,677\) & 0.29 & 0.36 \\
\(\{8,8\}\)&\([[25944,12974,8]]\)&\(38,916\)&\(487\)&68--80&\(52,223\)&\(41,186\) &\(52,210\) & \(41,120\) & 0.18 & 0.37 \\
\bottomrule
\end{tabular}
\end{table*}

Table~\ref{tab:routing-instances} summarizes both the topology-aware reference layouts and the variation across four additional independent topology-aware initializations for each of the four codes (20 runs in total). Every run successfully satisfies all hardware size and topological geometry constraints. To quantify initialization sensitivity, we define the relative range for a communication metric \(X\in\{R,K\}\) as:
\begin{equation}
\delta_X=100\frac{X_{\max}-X_{\min}}{\operatorname{median}(X)}\ \%.
\label{eq:routing-variation}
\end{equation}

\begin{figure*}[t]
\centering
\includegraphics[width=\textwidth]{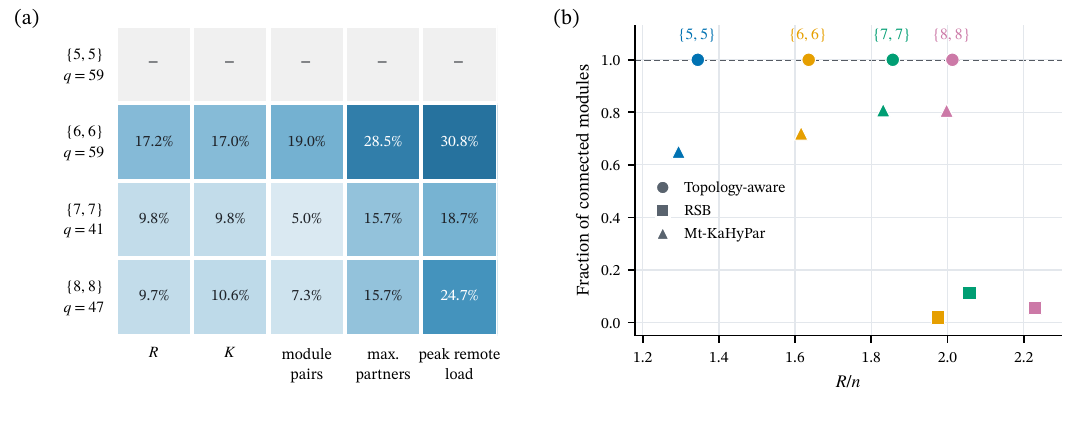}
\caption{Communication cost and connectivity constraints for the four code instances in Table~\ref{tab:routing-instances}. (a) Percentage reduction of the reference topology-aware layout relative to RSB, calculated for a metric $X$ as \(100(X_{\mathrm{RSB}}-X_{\mathrm{top}})/X_{\mathrm{RSB}}\). Rows identify codes by tiling, while columns represent the five communication metrics: \(R\), \(K\), the number of communicating module pairs, the maximum number of communication partners, and the maximum remote-interaction count. Dashes mark the \(\{5,5\},q=59\) case, for which RSB produced no valid capacity assignment in five attempted initializations. (b) Remote interactions per data qubit, \(R/n\), and the fraction of modules with connected induced interaction graphs. Colors identify tilings and symbols identify different compilers. The unavailable RSB assignment has no point. All tested topology-aware layouts satisfy all capacity and geometric constraints, whereas none of the tested RSB or Mt-KaHyPar assignments satisfies all these requirements.}
\label{fig:routing-benchmark}
\end{figure*}

For the \(\{5,5\}\) code, the topology-aware reference layout partitions the code into 1,155 modules, each containing 77 to 80 qubits, yielding \(R=69,021\) and \(K=57,382\). Here, remote interactions account for \(33.6\%\) of the \(4n\) total interactions per syndrome round. In this specific case, RSB failed to produce a physical assignment that satisfies the capacity constraints across five distinct initializations of the face partitioning step; thus, we omit RSB from the percentage-reduction comparison for this code. We emphasize that these results reflect failure of the specific RSB conversion heuristic tested, rather than a proof that no feasible RSB-derived assignment can be obtained.
For the remaining three instances, the topology-aware protocol strictly outperforms the tested RSB baseline across all five communication metrics, as detailed in Fig.~\ref{fig:routing-benchmark}(a). The RSB assignments failed to meet the geometric constraints as only \(1.8\%\), \(11.2\%\), and \(5.5\%\) of the RSB modules yielded connected local interaction graphs for the \(\{6,6\}\), \(\{7,7\}\), and \(\{8,8\}\) codes, respectively. 

All tested Mt-KaHyPar assignments satisfied the capacity
constraints, but none satisfied all geometric constraints.
The connected-module fractions for the \(\{5,5\}\),
\(\{6,6\}\), \(\{7,7\}\), and \(\{8,8\}\) codes are
\(65.0\%\), \(72.1\%\), \(80.8\%\), and \(80.7\%\),
respectively. All modules in the \(\{5,5\}\) and
\(\{6,6\}\) assignments are planar, whereas the
\(\{7,7\}\) and \(\{8,8\}\) assignments each contain
six nonplanar modules.
The communication comparison shows tradeoffs. Mt-KaHyPar
achieves lower remote-interaction counts \(R\), but higher
check fragmentation \(K\) and a larger maximum number of
communication partners per module, for all four codes.
It produces fewer communicating module pairs for the
\(\{6,6\}\), \(\{7,7\}\), and \(\{8,8\}\) codes, but
more for \(\{5,5\}\). Its maximum remote-interaction count
per module is lower for \(\{5,5\}\), \(\{6,6\}\), and
\(\{8,8\}\), but slightly higher for \(\{7,7\}\). For example, the
\(\{5,5\}\) Mt-KaHyPar assignment gives \(R=66\,427\)
and \(K=59\,698\), compared with \(R=69\,021\) and
\(K=57\,382\) for the topology-aware reference layout, while still failing to satisfy all geometric constraints.

We conclude that none of the tested baseline assignments satisfies all hardware layout constraints. 
A visual summary of communication costs and connectivity is provided in Fig.~\ref{fig:routing-benchmark}(b). 
These benchmark results demonstrate that our topology-aware compiler offers a highly practical mechanism for modularizing large-scale hyperbolic surface codes into bounded planar modules that satisfy capacity and geometric constraints, while reducing communication overhead relative to the tested load-feasible RSB assignments.

\section{Memory performance with noisier long-range operations}
\label{sec:circuit}

The hardware assignments described in Secs.~\ref{sec:architectures} and~\ref{sec:compiler} determine which two-qubit operations require nonlocal interactions. We now
evaluate the sensitivity of hyperbolic surface code quantum memories to errors in these operations,
using the largest four members of the closed $\{5,5\}$ code family presented in
Table~\ref{tab:scaling-sequences}.
For each instance, we compare two fixed assignments of nonlocal gates: an
effective-modular implementation based on our topology-aware partitioning algorithm described in Sec.~\ref{sec:compiler} and
a monolithic implementation with native long-range connections. The hyperbolic surface codes, as well as their selected logical operator, ideal extraction circuit, and noise probabilities are otherwise held fixed. 
This isolates the effect of the long-range interactions  and their associated gate-error penalties within a common circuit-level noise model.

Each experiment performs \(d\) consecutive syndrome-extraction
rounds, where \(d\) is the homological code distance. For the simulated
\(\{5,5\}\) codes, each check involves five data qubits and requires
five CNOTs between its ancilla and those qubits. In each round,
the ancillas are prepared, the CNOTs are applied in five parallel
layers for the vertex checks followed by five for the face checks,
and the ancillas are measured. Within each layer, no qubit
participates in more than one CNOT. Since each data qubit
participates in two vertex checks and two face checks, a complete
round contains \(4n\) CNOTs, where \(n\) is the number of data qubits.
We perform $Z$-memory experiments and measure the probability $p_L$ that decoding fails to recover one selected logical $Z$ operator after the complete experiment. 
We use a stochastic circuit-level noise model inspired by
SI1000, designed to be representative of the underlying error probabilities in contemporary superconducting quantum chips~\cite{GidneyEtAl2021SI1000}. A local CNOT is followed by two-qubit depolarizing noise with probability $p$. To account for the degraded fidelity of nonlocal gates, a nonlocal CNOT has error probability $\alpha p$, with $\alpha\in\{1,2,3\}$. Reset and measurement errors occur with probabilities $2p$ and $5p$, respectively. Idle qubits experience depolarizing noise with probability $p/10$ during ordinary gate layers and $2p$ while waiting through reset or measurement layers.

We use Stim to generate the circuits and the detector error
models~\cite{Gidney2021}, while decoding is done using correlated minimum-weight perfect matching implemented in PyMatching~\cite{Higgott2022}. 
For the selected logical $Z$-memory circuits, Stim's graphlike-error search gives circuit-fault distances $D=4,5,7,8$, compared with homological distances $d=6,8,10,12$.
These reduced distances characterize fault propagation through the specified measurement schedule; they do not alter the homological code parameters used in Secs.~\ref{sec:construction} and~\ref{sec:family}. In particular, a different noise model or measurement schedule might result in a different circuit-fault distance, while the homological distance is agnostic to the underlying error model and only depends on the code's structure.

For the modular implementation, we use the topology-aware disk
partitioning algorithm outlined in Sec.~\ref{sec:compiler}. A CNOT is nonlocal precisely when its data and ancilla
qubits belong to different modules. For the largest instance, 51,330 data
qubits and 41,064 check ancillas occupy 1,155 modules; 69,021 of the
205,320 CNOTs in each round are nonlocal. The inter-module
fraction remains close to one-third across the four sizes.

For the monolithic implementation, we represent the finite
\(\{5,5\}\) code in the Poincaré disk model. Vertex-check ancillas are placed at the lattice vertices, face-check ancillas at the geometric
centres of faces, and data qubits at edge midpoints.
A CNOT is classified as local when it connects an ancilla to a data qubit on an incident edge in this representation. 
A CNOT is classified as long-range when its data and ancilla qubits positions are connected through a periodic identification rather than a local incidence.

\begin{table*}[t]
\caption{Code parameters and interaction assignments for the four
simulated $\{5,5\}$ instances. Each round contains $4n$ CNOTs. Modular remote
counts are inter-module data--ancilla interactions; monolithic long-range
counts refer to nonlocal incidences between data--ancilla qubits.}
\label{tab:circuit-codes}
\centering\small
\setlength{\tabcolsep}{5pt}
\begin{tabular}{ccccccc}
\toprule
$q$ & $[[n,k,d]]$ &  Rounds & Modules & Modular remote &
Monolithic long-range & Fractions (mod./mono.)\\
\midrule
11 & $[[330,68,6]]$  & 6 & 8 & 431 & 405 & $32.65\%/30.68\%$\\
19 & $[[1710,344,8]]$  & 8 & 39 & 2,266 & 2,120 & $33.13\%/30.99\%$\\
31 & $[[7440,1490,10]]$  & 10 & 168 & 9,895 & 9,327 & $33.25\%/31.34\%$\\
59 & $[[51330,10268,12]]$  & 12 & 1,155 & 69,021 & 64,958 & $33.62\%/31.64\%$\\
\bottomrule
\end{tabular}
\end{table*}

Table~\ref{tab:circuit-codes} summarizes both assignments. At $\alpha=1$,
the modular and monolithic constructions give identical noisy circuits
for a given code, and therefore the same
underlying logical error probability; both yielding an estimated threshold of \(\approx 0.22\% \). For $\alpha>1$, both the locations and
the number of penalized gates differ between architectures. Any observed
difference in performance thus pertains to these discrepancies.

\begin{figure*}[t]
\centering
\includegraphics[width=\textwidth]{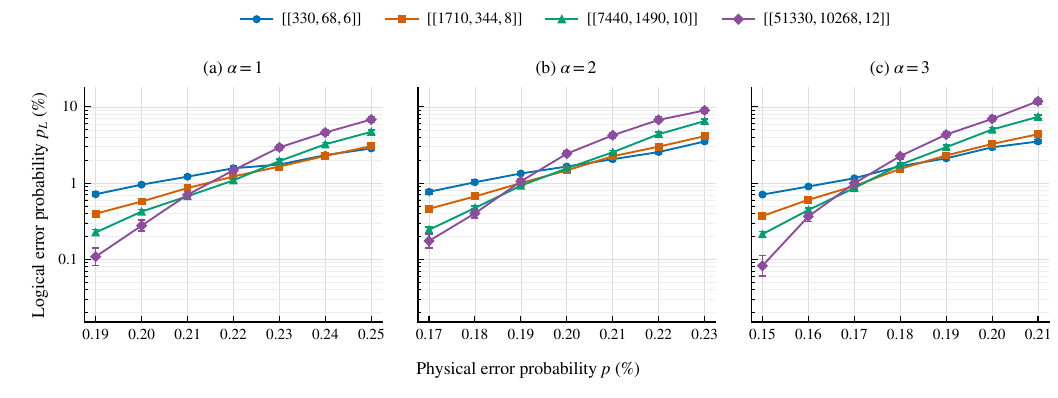}
\caption{Logical error probability $p_L$ for a selected logical $Z$ memory
in the effective-modular $\{5,5\}$ architecture, after $d$ syndrome rounds.
Panels (a)--(c) apply $\alpha=1,2,3$ to inter-module CNOT error
probabilities. Each curve contains seven measured points. Error bars are nominal pointwise $95\%$ Wilson intervals. Finite-size threshold estimates are $0.22\%$, $0.19\%$, and $0.17\%$, respectively.}
\label{fig:threshold-curves}
\end{figure*}

\begin{figure*}[t]
\centering
\includegraphics[width=\textwidth]{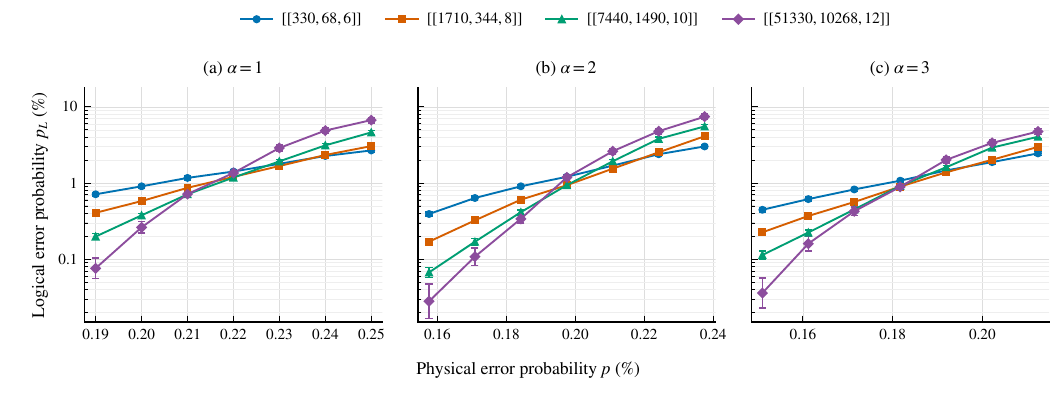}
\caption{Logical error probability for a selected logical $Z$ memory in
the monolithic $\{5,5\}$ architecture, with $\alpha$ applied to native
long-range CNOT errors. Code labels, syndrome rounds, marker conventions,
and error bars follow Fig.~\ref{fig:threshold-curves}. The finite-size threshold estimates for $\alpha=1,2,3$ are $0.22\%$,
$0.20\%$, and $0.18\%$.}
\label{fig:monolithic-pilot}
\label{fig:monolithic-threshold-curves}
\end{figure*}

For the Monte Carlo simulations, the codes use the following production rule: at least
$10,000$ shots per point, followed by stopping at $1,000$ logical failures or
$250,000$ shots. 
Figures~\ref{fig:threshold-curves} and~\ref{fig:monolithic-pilot} show seven
points per curve, shared by all four code sizes within each panel. 
To appreciate the large-scale of the simulations, note that the displayed modular data comprise 84 points, $6,501,136$ shots, and $70,338$ logical failures. The displayed
monolithic data likewise comprise 84 points, with $8,099,548$ shots and
65\,963 failures.
Both axes express error probabilities in percent, and the logical error probability is shown on a logarithmic scale, while error bars represent $95\%$ Wilson intervals.
For each architecture and \(\alpha\), we estimate the finite-size threshold by taking the median of the six pairwise intersection probabilities of the four code-size curves. Each intersection is obtained by linearly interpolating the difference between the measured logical error probabilities at adjacent physical-error probabilities.

In both architectures, the largest code has a lower logical error
probability than the smaller codes at the low-noise end of the sampled
range, but a higher probability at the high-noise end. This behaviour
persists for all three $\alpha$ values and is consistent with finite-size
threshold behavior. Increasing the nonlocal-gate penalty shifts the
crossing region toward lower physical error probabilities, demonstrating
the sensitivity of the memory experiment to the fidelity of the designated
nonlocal two-qubit gates.

The central result of this section is a circuit-level comparison of finite-size threshold estimates for two explicit hardware assignments: both exhibit $\approx0.22\%$ estimated thresholds at $\alpha=1$, while tripling the nonlocal-gate error probability reduces the estimated thresholds to $\approx0.17\%$ and $\approx0.18\%$ for the modular and monolithic architectures, respectively. These results agree with our understanding that the increased number of non-local interactions for the modular architecture compared with the with monolithic architecture, as shown in Table~\ref{tab:circuit-codes}, shall result in lower threshold estimates.

\section{Conclusion and discussion}
\label{sec:discussion}

Hyperbolic surface codes offer a compelling combination of high
encoding rates, growing code distances, low check weights, and a
geometric structure that organizes both logical information and
physical interactions. 
In this work, we operationalize this theoretical potential
through an exhaustive finite code search, hardware-aware layout
design, and circuit-level memory simulations that quantify the
cost of noisy long-range interactions. 
Within the specified self-dual
prime-field construction, for \(p\in\{5,6,7,8\}\) and \(q<60\), we identified a maximum code efficiency of approximately
\(33.34\) for the weight-six \([[51330,17112,10]]\) code. We also
presented a weight-five \([[51330,10268,12]]\) code with code
efficiency approximately \(28.81\). Crucially, we demonstrated that optimizing periodic surface
identifications can double the code distance at fixed physical-qubit
count, encoding rate, and check weight, increasing code efficiency
by a factor of four. We also presented weight-five and
weight-seven sequences that exhibit finite-size efficiency growth
consistent with the optimal logarithmic-squared scaling allowed
by Delfosse's bound; see Table~\ref{tab:scaling-sequences}.
We demonstrated that topological code parameters and physical hardware layouts provide complementary avenues for optimization. 
While the local hyperbolic tiling dictates the check weight and asymptotic encoding rate, the global periodic identifications govern the shortest
nontrivial logical cycles and hence the code distance. 

We also presented a topology-aware modular compiler that translates the underlying hyperbolic surface code into bounded-size planar modules while preserving the code structure. All 20 tested topology-aware layouts
satisfied the capacity and geometric constraints. For the three
benchmark codes with load-feasible RSB assignments, the topology-aware reference layouts reduced all five communication metrics. 
We also presented a monolithic construction that provided a complementary assignment of local and long-range data--ancilla interactions for a single-chip implementation.

We evaluated these assignments through circuit-level Monte Carlo
simulations of a selected logical \(Z\) memory across homological
distances \(d\in\{6,8,10,12\}\). Tripling the error probability of
designated long-range gates lowers the descriptive finite-size
threshold estimate from approximately \(0.22\%\) to \(0.17\%\)
in the modular architecture and from \(0.22\%\) to \(0.18\%\)
in the monolithic architecture.
Using the unrounded estimates, these reductions are approximately \(22.5\%\) and \(17.0\%\), respectively.
Both implementations retain finite-size threshold behavior under this increased long-range noise.

Beyond memory preservation, a practical large-scale fault-tolerant device requires constant-overhead gadgets that enable implementing a universal set of gates  \cite{zhu2020instantaneous, Lavasani2019universallogical}.
Within the surface-code paradigm, techniques such as lattice surgery, dynamical boundary deformations, and homomorphic logical measurements offer established pathways to logical control~\cite{BreuckmannEtAl2017,HuangEtAl2023}. 
Alternatively, in measurement-based quantum computing, hyperbolic cluster states have been demonstrated to exhibit favorable encoding efficiency while achieving circuit-level thresholds comparable to their Euclidean, vanishing-rate counterparts \cite{MahmoudEtAl2026Cluster}. 
These advances motivate exploring measurement-based gadgets for logical
operations.
A critical next step for this architectural roadmap is then the integration of these logical operations directly into the modular network or monolithic routing layers developed herein.

Furthermore, quantum geometry offers a further direction for investigating
the physical modes used to implement hyperbolic connectivity.
For isolated bands in Euclidean periodic systems, the Brillouin-zone average of the trace of the quantum metric gives lower bound on, the spread of Wannier functions~\cite{MarzariVanderbilt1997,MarzariEtAl2012}.
Our formulation of quantum geometry in hyperbolic band theory
provides a starting point for exploring related questions in
hyperbolic systems~\cite{mahmoud2026quantum}. Establishing this connection
could help determine whether quantum geometry provides useful
design criteria for mode localization and crosstalk in superconducting
devices.

A related question is whether quantum geometry can guide the control of long-range interactions. For pure states, the quantum Fisher information matrix is proportional to the quantum metric.
This relation motivates investigating control protocols for adiabatic gates and state transport in hyperbolic networks~\cite{ProvostVallee1980,BraunsteinCaves1994,ZanardiEtAl2007}.

Beyond superconducting devices,
photonic~\cite{WalsheEtAl2025,chen2026fusion},
neutral-atom~\cite{BluvsteinEtAl2023,xu2024constant},
and trapped-ion~\cite{HongEtAl2024,pino2021demonstration}
architectures offer routes to the nonlocal connectivity required by hyperbolic surface codes. This motivates developing layouts and syndrome-extraction circuits tailored to each platform, and evaluating their memory performance through circuit-level simulations for comparison with the superconducting models studied here.

We envision that the results presented in this work, alongside our previous results \cite{MahmoudAliRayan2026, XuEtAl2025, MahmoudEtAl2026Cluster}, establish a viable foundation for the realization of large-scale and efficient quantum memories based on hyperbolic surface codes.

\section*{Acknowledgments}
We gratefully acknowledge the developers of the \texttt{hypertiling} package~\cite{schrauth2023hypertiling}, which was utilized to generate the hyperbolic lattice depicted in Fig.~\ref{fig1:overview}(a). A.A.M. extends his gratitude to the organizers of the 2026 AlgoLab Summer School for facilitating an environment of stimulating discussion that helped shape this work. In particular, A.A.M. thanks Stefanos Kourtis and Shruti Puri for their insightful discussions on the theoretical bounds of qLDPC codes, as well as Xicheng (Christopher) Xu and Noah Gorgichuk for valuable conversations regarding superconducting circuit architectures.
S.R. and A.A.M. were supported by the NSERC Discovery Grant and PIMS Site Director Support Grant (both of S.R.).

\section*{Data availability}
The data and source code used in this study are available from the corresponding authors upon reasonable request.

\bibliography{references}
\end{document}